\documentclass[article,superscriptaddress]{revtex4}
\usepackage{mathrsfs}
\usepackage{etoolbox} 
\usepackage{lipsum} 
\usepackage{multirow}
\usepackage{dsfont}
\usepackage{amsfonts}
\usepackage{amsmath}
\usepackage[capitalize]{cleveref}
\usepackage{amssymb}
\usepackage{graphicx}
\usepackage{bbold}
\usepackage{dcolumn}
\usepackage{bm}
\usepackage{xcolor}
\usepackage{mathrsfs}
\usepackage[utf8]{inputenc}
\usepackage{amsmath}
\mathchardef\arr="017E 
\newcommand\undervec[1]{\setbox0=\hbox{$#1$}\lower2ex\hbox to 0pt{\hbox to \wd0{\hss$\arr\;$\hss}\hss}\box0}

\begin{document}
\title{Tunable mode entanglement: topological Su-Schrieffer-Heeger (SSH) chain with an embedded Aharonov-Bohm quantum ring}
\author{A. P. Garrido}
\affiliation{Departamento de F\'{\i}sica, Facultad de Ciencias, Universidad Cat\'olica del Norte,  Angamos 0610, Antofagasta, Chile}
\affiliation{Departamento de F\'{\i}sica, Universidad T\'{e}cnica Federico Santa Mar\'{\i}a, Valpara\'{\i}so, Chile}
\author{A. R. Plastino}
\affiliation{CeBio y Departamento de Ciencias B\'asicas,
      Universidad Nacional del Noroeste de la Prov\'\i ncia de Buenos Aires,
      UNNOBA, Conicet, Roque Saenz Pe\~na 456, Junin, Argentina}
\author{M. L. Ladr\'on de Guevara}
\affiliation{Departamento de F\'{\i}sica, Facultad de Ciencias, Universidad Cat\'olica del Norte,  Angamos 0610, Antofagasta, Chile}
\author{V. M. Apel}\email{vapel@ucn.cl}
\affiliation{Departamento de F\'{\i}sica, Facultad de Ciencias, Universidad Cat\'olica del Norte,  Angamos 0610, Antofagasta, Chile}
\date{\today}
\begin{abstract}
We investigate a variant of the SSH model consisting of an SSH chain with an embedded Aharonov-Bohm quantum ring. The embedded ring gives rise to domain wall states whose energy levels are in the band gap. The dependence of some of the states on the ring's magnetic flux provides a mechanism to control the entanglement between the two SSH edge fermion modes. The concurrence between the edge modes depends periodically on the magnetic flux. In the deep topological zone, for appropriate values of the flux, the concurrence reaches the extreme values of $0$ and $1$, corresponding to vanishing entanglement and to maximum entanglement. Therefore, the QR-SSH system offers a tuning mechanism that can be used either as an entanglement modulator or as an entanglement switch.
\end{abstract}
 \maketitle
\section{Introduction}

The SSH model, which describes the polyacetylene molecule, constitutes a paradigmatic example of a model quantum system admitting non-trivial topological  properties \cite{SSH}. The quintessential feature of this kind of systems is that they can hold gapless states at the edges that are immune to weak or moderated disorder. As an example we can mention the  quantum Hall bridge which, in the appropriate regime, while possessing robust metallic states at its edges is insulator at the bulk. The conducting states occur within the so called topological phase. In contrast to the topological phases, there exist also trivial phases where such states do not arise. Bearing in mind that the edges of a system have one dimension less than the bulk dimension, it is plain that for $1$D topological systems the edges states are zero dimensional, and therefore do not conduct. The topological states are then known as edge zero modes.
The different phases in a topological system can be explained in terms of topological properties of a gapped Hamiltonian. It is said that two Hamiltonians belong to the same topological class if, in the parameter space where they live, there is a continuous path along which it is possible to transform one Hamiltonian into the other without closing the gap.
There is a subcategory within the topological systems called symmetry protected topological (SPT) systems, in which certain symmetry properties of the Hamiltonian are preserved during the transformation \cite{classification}. The SSH model falls into such category.

The $1$D SSH system is governed by a one dimensional tight-binding Hamiltonian with  periodically alternating hoppings between neighbor sites \cite{SSH,SSH2}, that describes the transport of spinless fermions in a dimerized one-dimensional lattice. The difference between the values of the alternate hoppings is usually expressed in terms of an asymmetry parameter $\lambda$, adopting values in the range $[-1,1]$. This parameter determines  the  two phases of the system: the trivial phase corresponds to $\lambda <0$  and the topological one to $\lambda >0$.

Although originally proposed in the context of solid state physics,
 SSH-like models have recently been the focus of considerable attention in various other fields, that include quantum optics and quantum information theory. The increasing attention is due, among other things, to the  appealing variety of non-trivial phenomena that can take place in models based on SSH chains. For instance, it
 has been pointed out that the non-local nature of topological phases should be reflected in purely quantum mechanical correlations, such as quantum entanglement \cite{EntanglementSpectrumFidkowski,TopologicalEntanglementPointOfView,qctopologycal,LimaQuantumEntanglement,DisconnectedPartitionEntanglement,FermionicEntanglementNegativity,QuantumIformationMeetsCondensedMatter}.
Regarding this point, Cho and Kim \cite{Cho} discovered that mode entanglement between two first neighbor sites of an SSH chain exhibits  critical behaviour at the phase transition point ($\lambda=0$). Monkman \cite{Monkman2020} showed that the topological edge states
in a three leg SSH ladder can be a source of genuine non-bipartite many-body entanglement that can be transferred to a quantum register. These studies are related to another interesting phenomenon that deserves investigation: the appearance of domain walls in the SSH chains. In this regard, Huda {\em et al.}  recently reported the experimental realization of  dimer and trimer topological chains with domain wall structures precisely controlled  by atomic manipulation \cite{HudaTuneable}.

In the present work we investigate an SSH-based model, paying special attention to two  of the phenomena mentioned above: quantum entanglement and the existence of domain walls.
 We consider the SSH-based model schematically depicted in Fig. \ref{Fig1}, which consists of an SSH chain with an embedded  Aharonov-Bohm quantum ring (QR). The motivation for including the ring is to explore how the concomitant magnetic flux affects the system's topological states, and the mode entanglement features of the system's ground state. In addition to the topological edge modes, we have found that the ring embedded in the SSH chain configures a domain wall where two different kind of topological states are supported, according to the value adopted by the chain's asymmetry parameter. One of the states has zero energy, independently of the magnetic flux, while the other, which is  within the gap, has an energy tunable by the flux.  The dependence of the topological state on the ring's magnetic flux provides a mechanism to control the mode entanglement between the edge modes. In the deep topological zone ($\lambda\approx 1$), as the flux is varied,  the concurrence  has a discontinuous jump from $0$ to its maximal value $1$. In addition, there are critical values of  both the flux and the asymmetry parameter for which sudden births or deaths of entanglement are observed.

The article is organized as follows.  The model to be investigated in the present work
is described in section \ref{model}. Some necessary background material on the phenomenon of entanglement between fermion modes, including relevant technical details  concerning reduced density operators in the context of identical fermions,  are reviewed in section \ref{fermiback}.
Our main results are presented in section \ref{results}, where the system's energy spectrum and topological states are determined, and the states' entanglement features are investigated. Finally, some conclusions are drawn in section \ref{conclusions}.
\section{The Model} \label{model}
\begin{figure}[h]
\includegraphics[scale=0.25]{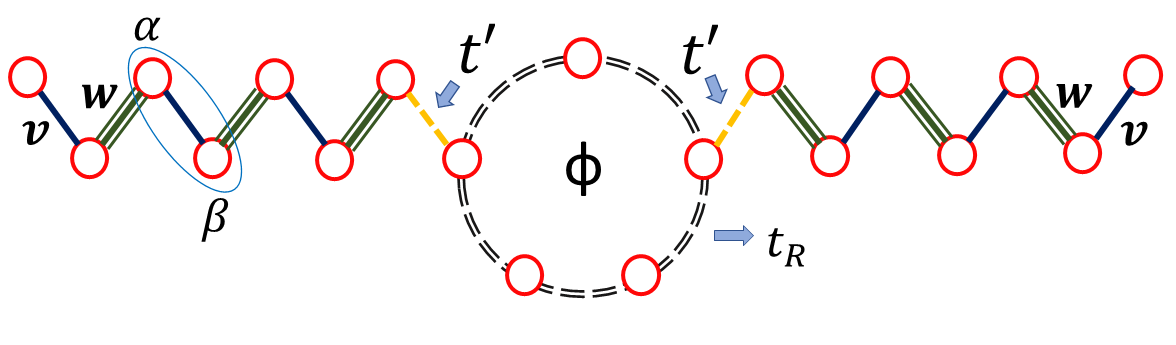}
\caption{ A Su-Schrieffer-Heeger (SSH) chain with an embedded Aharonov-Bohm quantum ring (QR). The SSH chain comprises two branches. Each of them is composed of cells containing two sites: one site of type $\alpha$ and one site of type $\beta$.}\label{Fig1}
\end{figure}

The system under consideration consists of a SSH chain with an embedded Aharonov-Bohm ring in the form described in Fig. \ref{Fig1}.
Each branch of the SSH chain is composed of cells  containing two sites: one of type $\alpha$ and one of type $\beta$. The intra-cell and inter-cell hoppings  are $v$ and $w$, respectively. We denote
by $t_R$ the hoping between the sites in the ring, and by $t^\prime$ the chain-ring hoping. The Hamiltonian of the system can be written as
\begin{equation} \label{hamiltonian}
H= H_{SSH} + H_{QR} + H_{QR-SSH}.
\end{equation}
The term $H_{SSH}= H_{SSH}^L+ H_{SSH}^R$ describes the Hamiltonian of the left and right branches of the SSH chain, with
\begin{eqnarray} \label{hssh}
H_{SSH}^{L(R)}&=&  \sum_{i \in L(R)}  ( v c_{\beta,i}^\dag c_{\alpha,i} +  \mbox{h.c.})     \nonumber \\
&  &+  \sum_{i \in L(R)} (w c_{\alpha,i+1}^\dag c_{\beta,i} + \mbox{h.c.} ), \label{HSSHLR}
\end{eqnarray}
where the index $i$ corresponds to the $i$-th pair in the left (right) chain and $c_{\alpha(\beta)}$ denotes the spinless fermion operator in a $\alpha(\beta)$-type site. The intra and inter cell hopings, $v$ and $w$, can be expressed in terms of the asymmetry parameter $\lambda$ ($-1 \leq \lambda \leq  1 $) as
\begin{equation}
v=-t(1-\lambda) \; \mbox{ and } \; w=-t(1+\lambda),
\end{equation}
where $t$ has energy dimensions. $H_{QR}$ is the Hamiltonian of the quantum ring
\begin{equation} \label{hr}
H_{QR}= \sum_{j=1}^n t_R ( e^{i\varphi} d_{j}^\dag d_{j+1} + \mbox{h.c.}),
\end{equation}
where $n$ is the number of sites in the ring, $t_R$ the hoping between sites, $d_{j}$ is the spinless fermion operator on site $j$ of the ring, and $\varphi=2\pi \phi/\phi_0$ is the Aaharonov-Bohm phase, which is expressed in terms of the ring's magnetic flux $\phi$ and
the flux quantum $\phi_0=h/e$. Finally, $H_{R-SSH}$ is the Hamiltonian describing the coupling between the chains and quantum ring
\begin{equation} \label{hr-ssh}
H_{R-SSH} =  t^\prime (d_{L}^\dag c_{\alpha,N_{SL}} +  c^\dag _{\alpha,1_R} d_{R} ) + \mbox{h.c.},
\end{equation}
where $t^\prime$ is the chain-ring hoping and $d_{L(R)}$ destroys an electron in the site in the ring connected to the left(right) SSH chain.
\section{Reduced density matrices in the modes representation and
entanglement between fermionic modes}\label{fermiback}
In this section we discuss some background material on reduced density matrices within
the mode representation of fermionic systems, and on the quantitative evaluation
of entanglement between fermionic modes.
\subsection{Reduced density matrices in the fermionic modes representation}
 In order to calculate the amount of entanglement between the components of a subsystem of a quantum system, it is necessary first to obtain
the subsystem's reduced (also known as marginal) density matrix. For instance, to calculate the entanglement between two particular modes of a fermionic system, one needs first to calculate the reduced density matrix that jointly describes the two modes. When considering entanglement between modes in systems of identical particles, the natural framework to use is the occupation number representation \cite{UnifyingApproachReducedMatrix,Fermionic-mode}.
In the Appendix we describe the procedure to calculate the reduced density matrix in the occupation number (or fermionic modes) representation, which in addition will allow us to interpret  our results in the context of many identical particles.
In the modes representation for identical fermions, the natural basis of the many-body state-space (Fock space) consists of states characterized  by the occupation numbers  $n_s$ of each site $s$ ($1\leq  s \leq N_L$), where $N_L$ is the total number of sites in the system. The vacuum state $|\emptyset\rangle$ is the state with $n_s=0$ for all $s$. The basis of the Fock space contains the vacuum and states of the form $|I;n_{1},n_{2},\dots,n_{N_{L}}\rangle$, which are obtained by the action of the creation operators $c_{i}^{\dag}$ on $|\emptyset\rangle$,
\begin{subequations}
\begin{equation}
|I;n_{1},n_{2},\dots,n_{N_{L}}\rangle=\prod_{l=1}^{r}c^{\dag}_{i_{l}}|\emptyset \rangle,
\label{occ-creat}
\end{equation}
where $r$ is the number of fermions, the Fermi annihilation and creation operators obey $\{c_i,c^\dag_j\}=\delta_{ij}$, and we have adopted the convention,
\begin{equation}
\prod_{l=1}^{r}c^{\dag}_{i_{l}}=c^{\dag}_{i_{r}}\dots c^{\dag}_{i_{2}}c^{\dag}_{i_{1}}, \quad i_{1}<i_{2}<\dots<i_{r}. \label{prod_}
\end{equation}
\end{subequations}
\noindent
We have denoted by $I$ the multi-index $(i_1,i_2,\dots,i_{N_L})$.
In the state of Eq.(\ref{occ-creat}), the label $I$ indicates the occupied sites of the system, i.e.,  if $s$ ($1\leq s \leq N_L$) is equal to some index of $I$, then $n_s=1$,
otherwise, $n_{s}=0$.   Notice that, as follows from Eq. (\ref{prod_}), the indices in $I$ are defined in increasing order.

Given the anticommutative character of the fermionic operators, it is convenient to use the generalized Kronecker delta symbol, $\delta_{I}^{J}$, defined as,
\begin{equation} \label{generalkronecker}
 \delta^{J}_{I} = \begin{cases}
 1\textrm{ }\textrm{ }\textrm{ if }J  \textrm{ is an even permutation of }I\nonumber\\
    -1 \textrm{ if }J  \textrm{ is an odd permutation of }I\nonumber\\
  0 \textrm{ }\textrm{ }\textrm{ if }J  \textrm{ is not permutation of }I\nonumber
\end{cases}
\end{equation}
where $I=(i_{1},\dots,i_{f})$ and $J=(j_{1},\dots,j_{f})$ are two multi-indices. In this definition, the indices in $I$ or $J$ are not necessarily ordered.

Let $A|B$ be a partition of a lattice $L$, i.e., $L=A\cup B$, $A\cap B=\emptyset$, with $A=\{\alpha_{1},\dots,\alpha_{N_{A}}\}$ and
 $B=\{\beta_{1},\dots,\beta_{N_{B}}\}$,  where $N_{A}+N_{B}=N_{L}$.
As described in details in the Appendix, the reduced density matrix of the subpart $A$ in the fermionic modes representation is given by
\begin{subequations}
\begin{equation}
\rho_{A}=\sum_{J,J'}[\rho_{A}]_{J,J'}|J;n_{\alpha_{1}},\dots,n_{N_{A}}\rangle\langle J'; n_{\alpha_{1}},\dots,n_{N_{A}}|,
\end{equation}
\begin{equation}
[\rho_{A}]_{J,J'}=\sum_{I,I',K}\delta_{JK}^{I}\delta_{J'K}^{I'}[\rho]_{I,I'},
\end{equation}
\label{rho_A_fm}
\end{subequations}
where the labels $J$ or $J^\prime$  denote the sites occupied in the subpart A, $K$ denotes the sites occupied in subpart $B$, and $JK=(j_1<\dots <j_p, k_1< \dots < k_q)$, $p+q=r$, where $r$ is the number of occupied sites in all the system.
Notice that the indices in $JK$ are not necessarily ordered in increasing order, in spite that $J$ and $K$ are ordered.

An alternative procedure to obtain the reduced density matrix is based on the analytical procedure discussed by Peschel \cite{P2003} and by Cheung and Henley \cite{Cheong}, which is consistent with what was defined in Eqs.  (\ref{rho_A_fm}).  To calculate the reduced density matrix $\rho_{A}$ it is necessary to evaluate the correlation matrix $G$ defined by the elements $\langle c^{\dag}_{i}c_{j}\rangle$, where $i,j$ are the sites that belong to the subpart $A$. Once the correlation matrix $G$ is calculated, the reduced density matrix $\rho_{A}$ can be calculated using the formula,
\begin{eqnarray} \label{rho1}
\rho_{A}&=& \det(I - G) \exp{\left\{ \sum_{i j}\left[ \log G(I-G)^{-1} \right]_{ij} c_{i}^\dag c_{j}   \right\}},\nonumber\\
 &i,j&\in A.\label{MatDensCheong}
\end{eqnarray}
Although Eqs. (\ref{rho_A_fm}) and (\ref{MatDensCheong}) are equivalent, the latter is easier to implement in the calculation of reduced density matrices, and for  this reason we have used this to obtain our results.  As for Eq. (\ref{rho_A_fm}), its usefulness lies on the fact that it illustrates in a straightforward way the meaning of the reduced density matrix, and it highlights some technical aspect its computation when working in the modes representation.

\subsection{Entanglement between modes in fermionic systems} \label{mode-entanglement}
Entanglement is a key piece of quantum mechanical theory. This notion is central for understanding the physical properties of composite quantum systems, including systems consisting of identical particles. Considerable effort has been devoted to analyze and clarify the concept of entanglement in systems of identical particles
\cite{vedral,Compagno,shi-entanglement-identical-particles,zanardi,Wiseman,Bruschi,ZP2010,GR2015,MBVP2016}, and to apply it to the study of specific physical systems \cite{BMPSD2012,SN2015,IMV2015,MF2016,MF2018,TGR2018,TRCG2019,BCRM2019,BMV2020,C2021,FIMV2022,C2022,F2022}. In the present work, as already mentioned, we shall focus on the entanglement between modes. When addressing entanglement from the modes' perspective
each mode can be distinguished from the others, because each of them is associated to a specific lattice-site. Since the modes are spatially localized, the entanglement between modes considered here can be regarded as spatial entanglement. For the sake of clarity and completeness, it is worth to briefly discuss the notion of entanglement between modes. Following the Wiseman-Vaccaro argument \cite{Wiseman} let us consider, as a simple illustration of entanglement between modes, a four site system with only one electron in a superposition state
\begin{equation}
\frac{1}{\sqrt{2}}(|\varphi_{1}\rangle +|\varphi_{2}\rangle).
\end{equation}
This state doesn't seem to be entangled, since it
is a one particle state. In the occupation number
representation, however, this state is given by
\begin{equation} \label{belly}
\frac{1}{\sqrt{2}}(|1000\rangle+|0100\rangle),
\end{equation}
which is entangled. In fact, from the perspective of fermionic modes,
(\ref{belly}) is a Bell state.
The reverse situation can also occur. That is, a fermion state that in the particle representation looks like a Bell state, but in the mode representation is a separable state. For example, in our two spinless electron system in a four-site lattice, the state
\begin{equation}
 \frac{1}{\sqrt{2}}(|\varphi_{1}\rangle\otimes|\varphi_{3}\rangle-|\varphi_{3}\rangle\otimes|\varphi_{1}\rangle)
\end{equation}
looks non-separable. From the fermionic mode perspective, however,
this state is expressed as,
\begin{equation}
|1010\rangle
\end{equation}
which is a separable state. These simple examples correspond to pure states, but the concept of entanglement between fermionic modes can be straightforwardly extended to mixed states. The amount of entanglement between fermionic modes, both for pure and mixed states, can be calculated using standard procedures analogous to those used for systems consisting of distinguishable qubits. In this work, in order to calculate the entanglement between two modes jointly described  by a mixed state $\rho$, we shall use the concurrence $\mathcal{C(\rho)}$. To calculate $C(\rho)$, previously we require to specify an object called spin-flip superoperator $\tilde{\rho}$,
\begin{equation}\label{rhotilde}
\tilde{\rho} = (\sigma_y\otimes \sigma_y) \rho^* (\sigma_y\otimes \sigma_y),
\end{equation}
where $\rho$ is represented in the basis $\{|00\rangle, |01\rangle, |10\rangle, |11\rangle\}$, $*$ denotes complex conjugation, and  $\sigma_{y}$ is the Pauli matrix $\left(
                                                                           \begin{array}{cc}
                                                                             0 & -i \\
                                                                             i & 0 \\
                                                                           \end{array}
                                                                         \right)$,
($\sigma_{y}|0\rangle=i|1\rangle$ and $\sigma_{y}|1\rangle=-i|0\rangle$).
 The concurrence $\mathcal{C}$ is evaluated using Wootter's \cite{W1998} celebrated expression,
\begin{equation}\label{concurrencia-formula}
\mathcal{C}= \max\left\{0, \lambda_1- \lambda_2 -\lambda_3 -\lambda_4\right\},
\end{equation}
where $\lambda_{i}$ ($i=1,2,3,4$) are the eigenvalues, in decreasing order, of the operator $R=\sqrt{\sqrt{\rho}\tilde{\rho}\sqrt{\rho}}$.
The concurrence $\mathcal{C}$ can take values between $0$ and $1$, where $\mathcal{C}=0$ indicates absence of entanglement and $\mathcal{C}=1$ corresponds to a maximally entangled state.

\section{Topological States and Entanglement Between Modes}\label{results}
In the numerical simulations performed in this work, it is observed
that the properties of the system remain qualitatively invariant
when varying some of the model's parameters, such as the hoppings
$t_R$ and $t'$. In what follows, it is assumed that the
two SSH branches have the same number of sites, and that the ring has five sites. Also, the values of all the parameters appearing in
the Hamiltonian are express in units of $t$.
Our main results are presented in the following order: In subsection \ref{energy_spectrum} we analyze the effect of the quantum ring
on the energy levels of the system determining, in particular,  the levels'
dependence on the ring's magnetic flux.  In section \ref{entanglement_ss}
we study the amount of entanglement between fermionic modes, as measured by the concurrence evaluated on the marginal density matrix associated with a given pair of modes.
\subsection{Energy Spectrum and topological states} \label{energy_spectrum}
\begin{figure}[h]
\includegraphics[scale=0.72]{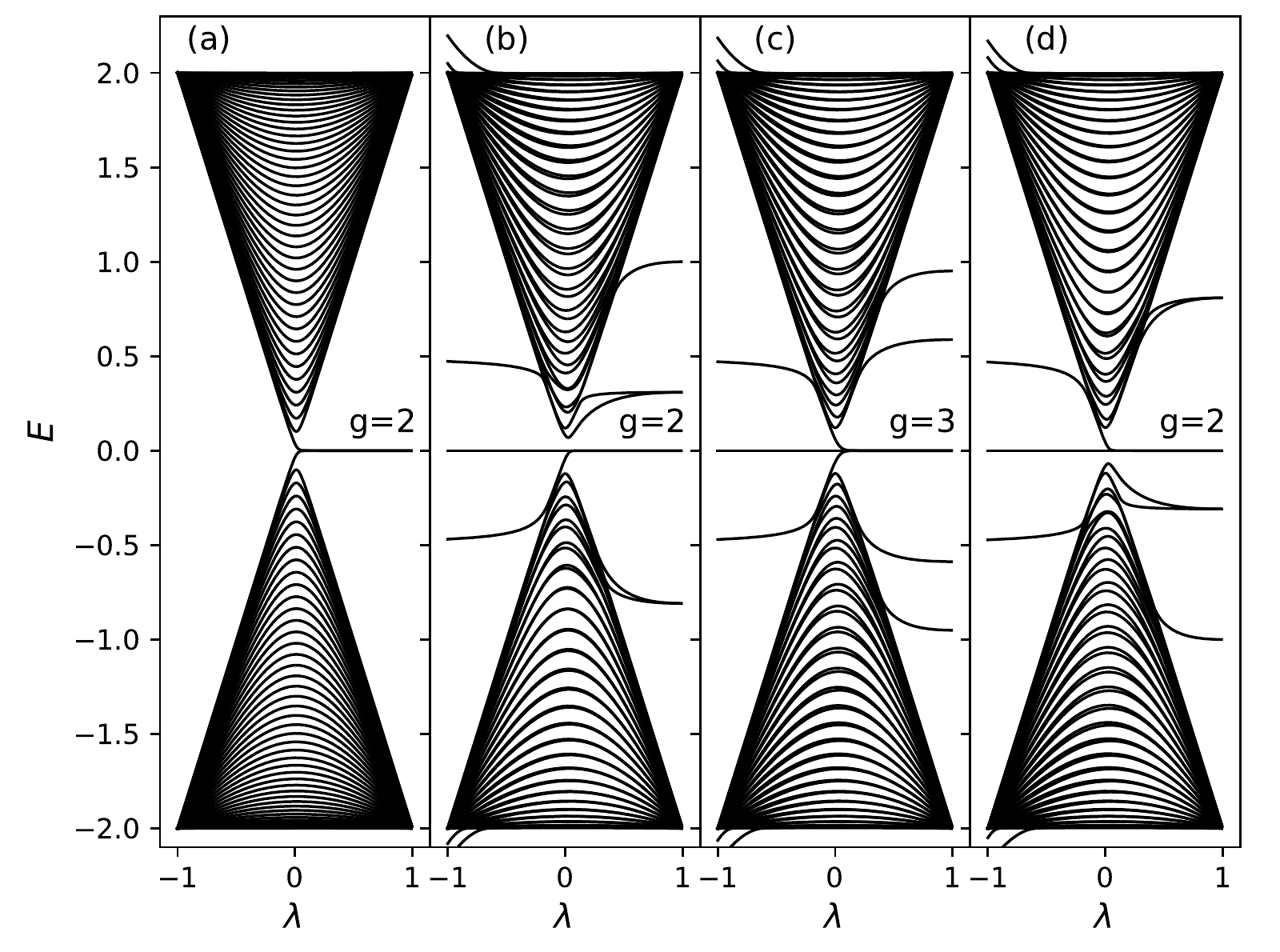}
\caption{Energy levels as a function of asymmetry parameter $\lambda$. (a) corresponds to standard SSH chain without QR with $N_T=90$ sites. (b), (c) and (d) correspond to SSH-QR system $N_T=97$ sites and $t_R=0.5t$.  The corresponding magnetic flux are (b) $\Phi/\Phi_0=0$, (c) $\Phi/\Phi_0=0.25$ and (d) $\Phi/\Phi_0=0.5$  }\label{Fig2}
\end{figure}

The energy spectrum provides essential information about the system's physical properties. In particular, it sheds lights on the presence of topological states.
For example, the spectrum of the standard SSH model presents a quasi-degenerate couple of levels around $E=0$, characterized by their robustness in the region of $\lambda >0$. The presence of such levels is associated with the existence of  topological states.
In the following analysis, as usually done when discussing the standard SSH chain, we refer to the region with $\lambda >0$ as the ``topological zone", and to the region with $\lambda<0$ as the ``trivial zone"  \cite{Janos}. In our model, however, the existence of topological states is not restricted to $\lambda>0$. Indeed, in the QR and its neighboring sites, topological states are observed for both $\lambda>0$ and $\lambda<0$.  Fig. \ref{Fig2} depicts the dependence of the system's energy levels on the asymmetry parameter $\lambda$. Fig. 2a corresponds
to a standard SSH chain, while Figures 2b, 2c, and 2d
correspond to the SSH-QR system with different values
of the ring's magnetic flux.
In the panels of Fig. \ref{Fig2} we specify the degeneration degree $g$ of the level $E=0$ in each zone. As shown in all the figures, there exist robust zero energy levels in the topological zone, but one such state is also observed in the trivial zone.
We find that the standard topological states at the free edges of the chains arise for any value of the magnetic flux (a two-fold degenerate state appearing in the region $\lambda>0$).
In the particular case of $\Phi/\Phi_0=0.25$ the degeneration in the topological zone is triple; to the standard edge states it is added a robust state localized at the ring and its neighbor sites. The robust level around $E=0$ for $\lambda <0$ indicates the presence of a non-expected topological state in this region,
which corresponds to a state located in some sites of the ring and its neighborhood.

\begin{figure}[ht]
\centering
\includegraphics[scale=0.4]{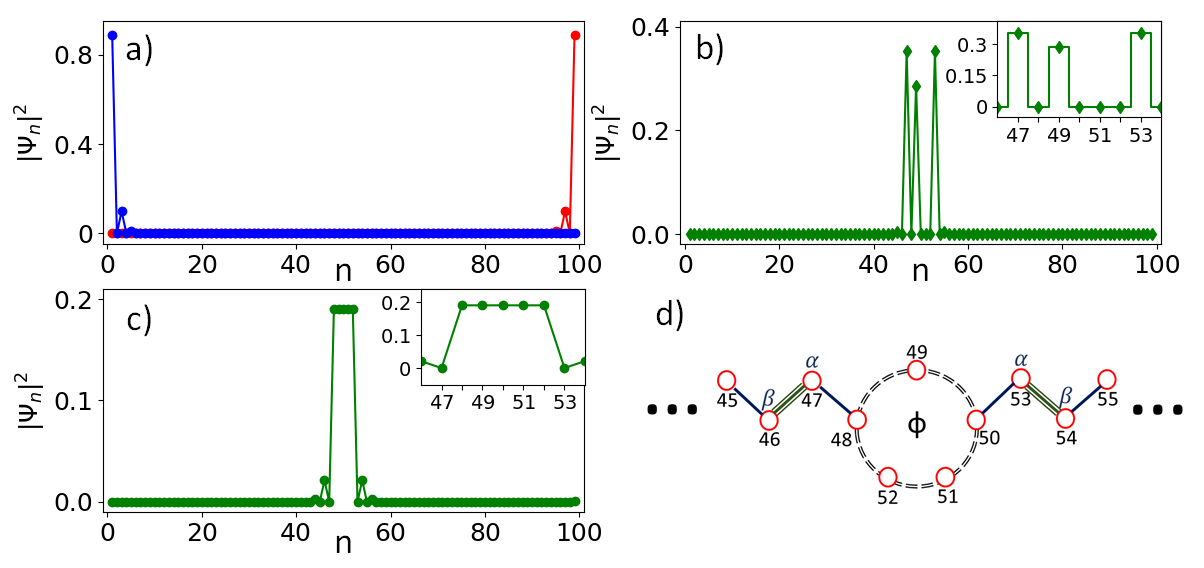}
\caption{Colour online. Panels (a), (b) and (c) show the probability as a function of the site position $n$ for three different of system's
eigenstates with energy $E/t\approx 0$ ($|E/t|\ll 1$). The panel (a) shows the topological edge states  for $\lambda=0.9$
and an arbitrary value of $\Phi$.
Panel (b) corresponds to the state with energy $E=0$ in the trivial zone, for $\lambda=-0.9$ and an arbitrary value of $\Phi$.
Panel (c) corresponds to an additional state in the topological zone
with $\Phi/\Phi_{0}=(l+1/4)$ ($l\in\mathds{Z}$) and $\lambda=0.9$. In
panel (d) we specify the correspondence of the values of $n$ their respective position in the system.}
\label{Fig3}
\end{figure}

\begin{figure}[h]
\includegraphics[scale=0.49]{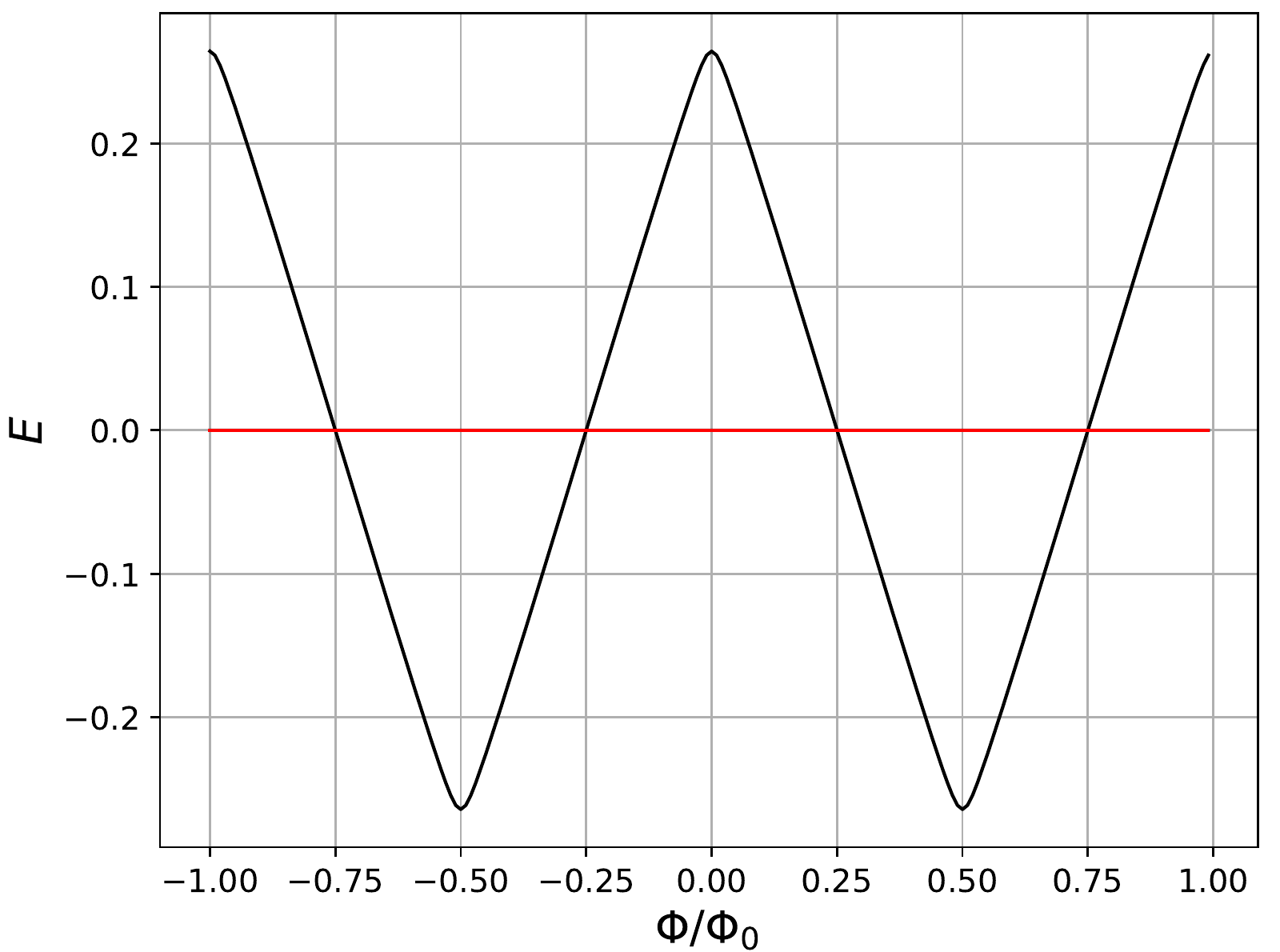}
\caption{Colour online. Energy of the QR confined states as a function of magnetic flux. Black  corresponds to QR confined state
in  the topological zone ($\lambda =0.4$), and constant energy level (red line) corresponds to localized QR state in the trivial zone ($\lambda<0$). The QR hoping is $t_{R}=0.5t$ }\label{Fig4}
\end{figure}

In Fig.  \ref{Fig3}, it is plotted  the probability that the electron is found at the $n$ site, where the features mentioned above are revealed.
In Fig.  \ref{Fig3}(a) we observe the standard topological edge states, which occur for $\lambda>0$ and any value of the magnetic flux.
Figs. \ref{Fig3}(b) and (c) give account of the two kinds of topological states confined in the QR region. Fig.  \ref{Fig3}(b) corresponds to the probability of the non degenerate eigenstate with energy $E=0$, for $\lambda<0$, which does not depend on the magnetic flux.
The electron's probability distribution localizes
just in one site of the ring and in some neighboring sites. This localization in one site of the ring explains why the energy of this state is insensitive to the magnetic flux. Let us recall that, for the Aharonov-Bohm effect to take place, it is necessary that the electric charge can circulate along a closed circuit around the magnetic field. For the state
under consideration, the probability that the electron transits along one of the ring's arms is zero, and, consequently, the magnetic field does not produce any observable effect in the system. In particular, the energy does not depend on the flux.
In the topological zone ($\lambda>0$), in addition to the two topological edge states, a third topological state arises, which has $E=0$ for $\Phi/\Phi_{0}=l+1/4$ ($l$ integer). The concomitant
probability distribution is uniform in the ring, as transpires from Fig. \ref{Fig3}(c).
In contrast with the other topological QR state, in this case the Aharonov-Bohm effect takes place, since the electron can circulate through both the ring's arms.
 Fig. \ref{Fig4} shows the energy as a function of the magnetic flux, for the two topological QR states. We can see that the energy of the state with uniform probability in the QR ($\lambda > 0$)
oscillates with the magnetic flux, while for the case of the state localized in only one QR site ($\lambda<0$), the energy is constant and equal to $0$.
This means that the former state supports persistent currents ($I_P=-dE/d\Phi$), while the latter does not. These topological states, which are localized in the ring and its surroundings, configure domain walls. It is worth emphasizing that the topological states having a uniform probability distribution in the ring do not necessarily have vanishing energy. In fact, the energy can be controlled by the intensity of the magnetic flux.
\subsection{Entanglement}  \label{entanglement_ss}
\begin{figure}[h]
\includegraphics[scale=0.4]{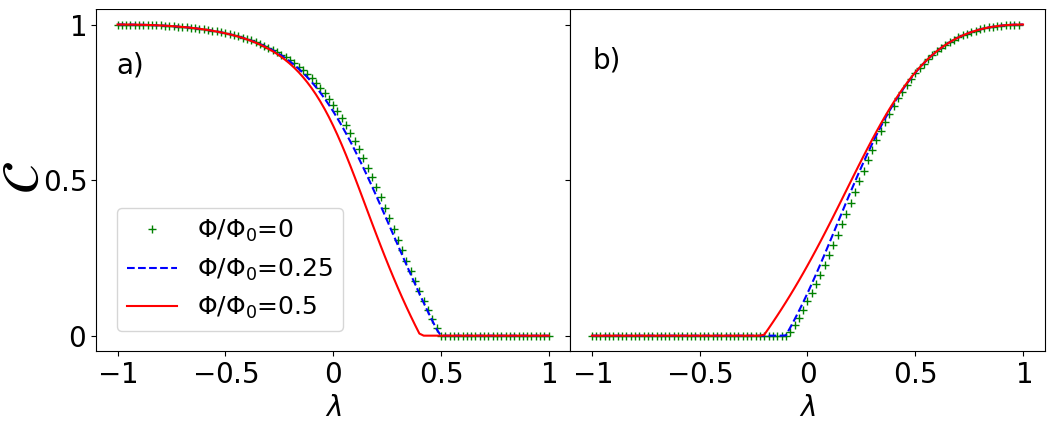}
\caption{Colour online. Concurrence between first neighbors as a function of $\lambda$ and $t_{R}=2t$. Panel a) corresponds to intra-cell sites, and panel (b) corresponds to
inter-cell sites. }\label{Fig5}
\end{figure}

In this section we investigate the amount of entanglement exhibited by particular couples of modes in the system, and analyze how entanglement is related to the topological phases. Specifically, we explore the concurrence between adjacent modes, as well as between the two edge modes. For relatively large systems (say, for systems with a total number of sites $N_T$ similar or greater than $20$) the numerical simulations are too costly in terms of computational time. Therefore, we have considered
$N_T= 19$ total sites: five sites in the quantum ring and  $7$ sites in each branch of the SSH chain. In spite of the size limitations, the results reported here strongly suggest that the observed effects can be extrapolated to systems of much larger size. Moreover, it is to be expected that some of the effects should be enhanced when considering larger systems. In all our computations we assumed that the system is near half-filling. That is, we assumed that the number of electrons is equal to $(N_T-1)/2$.
\\
\textit{Concurrence between nearest neighbors in the SSH chains.}
Here, we focus on the concurrence (see Eq. (\ref{concurrencia-formula})) between nearest neighbors in the SSH chains. The reduced density matrix corresponding to a pair of neighbour sites is obtained from Eq. (\ref{rho1}).
As shown in Fig. \ref{Fig5}, the concurrence between nearest neighbor sites of the chains can be different from zero. This is in agreement with the results reported by Cho and Kim \cite{Cho} for the standard SSH chain.
The panels (a) and (b) depict, for different values of the magnetic flux, the dependence on the parameter $\lambda$ of the concurrence $\mathcal{C}$ between nearest neighbors of the SSH chains. Panel (a) corresponds to next neighbours belonging to the same cell ($\alpha_n,\beta_n$), and panel (b) to next neighbours of adjacent cells ($\beta_n,\alpha_{n+1}$).
The qualitative behaviour of the concurrences are independent of the cell position.
Notice that $\mathcal{C}_{\alpha_n,\beta_n}(\lambda) \sim \mathcal{C}_{\beta_n,\alpha_{n+1}}(-\lambda)$, which follows from the fact that the
hoppings intra and inter-cell  satisfy the relation $v(\lambda)=w(-\lambda)$. In addition, the concurrence $\mathcal{C}_{\alpha_n,\beta_n}$
($\mathcal{C}_{\beta_n,\alpha_{n+1}}$) reaches its maximum $\mathcal{C}=1$ in the trivial (topological) zone, for $\lambda =-1$ ($\lambda=1$). To understand this behaviour, let us consider the limits $\lambda = \pm 1$. In the limit $\lambda=-1$ one has  $w=0$ and $v=-2t$, and  each branch of the SSH chain decomposes in decoupled bi-atomic molecules consisting of the sites ($\alpha_n,\beta_n$), that are in the
bonding states
\begin{eqnarray}\label{bellintracelda}
|\Psi_{n}(\lambda=-1)\rangle &=&\frac{1}{\sqrt{2}}(|\alpha_n\rangle+|\beta_{n+1}\rangle)\nonumber\\
 &= &\frac{1}{\sqrt{2}}(|1_{n}0_{n}\rangle+|0_{n}1_{n}\rangle).
\end{eqnarray}
Analogously, for $\lambda=1$ one has $v=0$ and $w=-2t$, and the SSH branches decompose in decoupled molecules consisting of the sites ($\alpha_n,\beta_n$), in the bonding states
\begin{eqnarray}\label{bellintercelda}
|\Psi_{n}(\lambda=1)\rangle &=&\frac{1}{\sqrt{2}}(|\beta_n\rangle+|\alpha_{n+1}\rangle)\nonumber\\
 &= &\frac{1}{\sqrt{2}}(|1_{n}0_{n+1}\rangle+|0_{n}1_{n+1}\rangle).
\end{eqnarray}
It is plain that both (\ref{bellintracelda}) and (\ref{bellintercelda}) are formally maximally entangled Bell-like states of two qubits.
When the value of $\lambda$ departs from $\pm 1$, each molecule interacts and becomes entangled with its neighbors, and the entanglement between next-neighbour sites decreases.
It is worth noticing that there are some critical values of $\lambda$ for which
entanglement between next-neighbour sites is completely suppressed \cite{Cho}, in a way reminiscent of the sudden death of entanglement predicted and observed in the context of quantum optics \cite{Davidovich}.

\begin{figure}[h]
\includegraphics[scale=0.33]{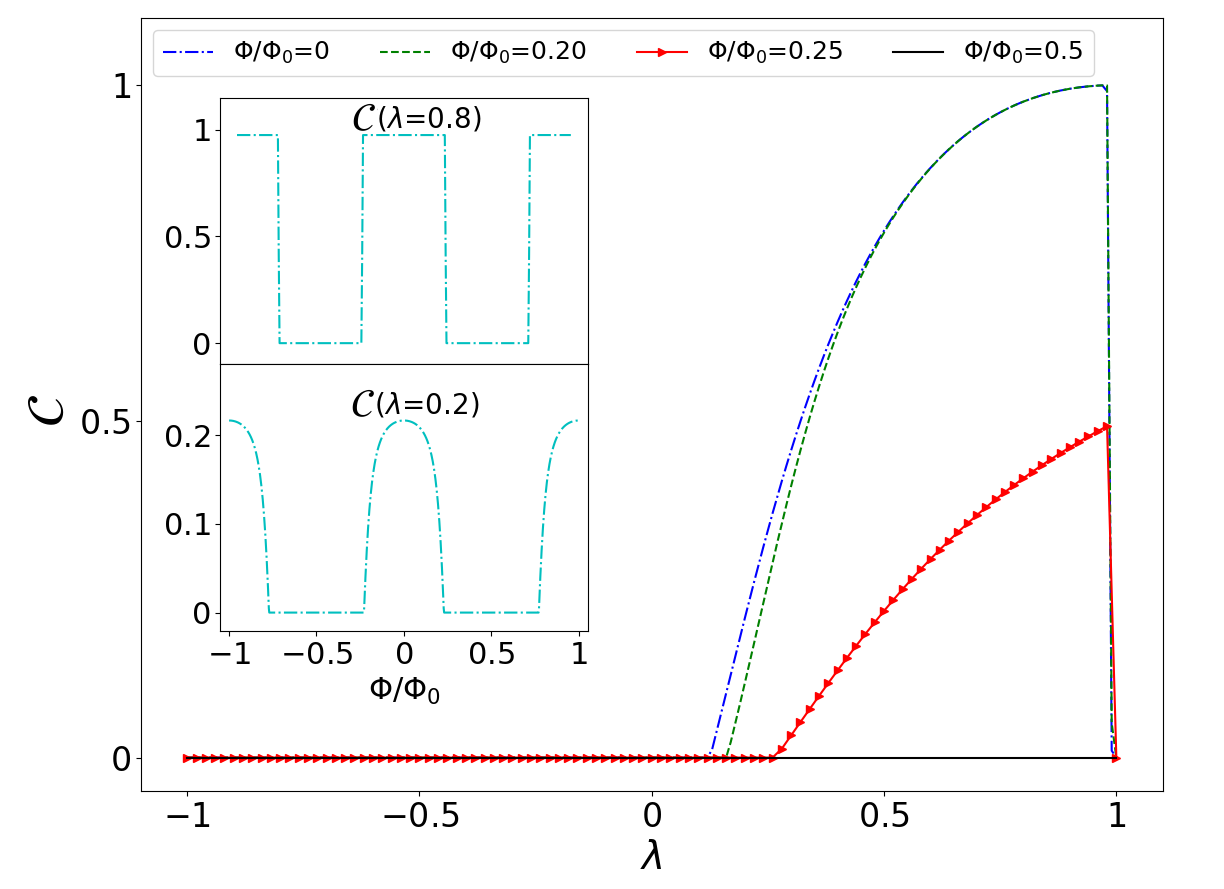}
\caption{Colour online. Edge-edge concurrence as a function of $\lambda$ for four values of magnetic fluxes and $t_{R}=2t$. The insets show the concurrence as a function of magnetic flux for two different values of $\lambda$.}\label{Fig6}
\end{figure}

\textit{Concurrence between edges sites.}
Here we present the central result of our work, the existence of entanglement between the edge sites of the system.
The behaviour of the concurrence $\mathcal{C}_{L,R}$ between the edge sites of the system is shown in Fig. \ref{Fig6}, as a function of the parameter $\lambda$,
for four different values of the magnetic flux ($\Phi/\Phi_{0}=0$, $0.2$, $0.25$ and  $0.5$ ).
In all cases the concurrence $\mathcal{C}_{L,R}$ vanishes in the trivial zone.
For $\Phi/\Phi_0=0.5$ the concurrence also vanishes in the topological zone, while for the remaining values of the ring's magnetic flux ($\Phi/\Phi_{0}=0$, $0.2$ and $0.25$) it departs from zero at a critical positive value $\lambda_{\Phi}$ of the asymmetry parameter, and then
increases monotonously with $\lambda$ in the interval $(\lambda_\Phi,1)$. In the limit $\lambda \to 1^-$  $\mathcal{C}_{L,R} \to 1$ for
$\Phi/\Phi_0= 0$ and $0.2$, and falls abruptly to zero at $\lambda=1$, when $v=0$ and the edge sites become disconnected from the rest of the SSH chains. For
$\Phi/\Phi_0= 0.25$ the concurrence behaves in a similar way, but its maximum value, reached in the limit $\lambda \to 1^-$, is $1/2$.
In the inset of the Fig \ref{Fig6}, we also show the dependence of the concurrence $\mathcal{C}_{L,R}$ on the
magnetic flux $\Phi$ for two fixed values of asymmetry parameter, $\lambda=0.8$ and $\lambda=0.2$. We observe that the concurrence between the edge sites is a symmetric and periodic function of the ring's magnetic flux $\Phi$, with period equal to $\Phi_0$.
In the deep topological  zone ($\lambda\approx 1$)  the concurrence is a discontinuous function of the flux,  with $\mathcal{C}_{L,R}=1$ for $|\Phi/\Phi_0| < 0.25$
and $\mathcal{C}_{L,R}=0$ for $0.25< | \Phi/\Phi_0| < 0.5$.

The entanglement between the edge modes is closely related to the occupation of the quasi-degenerate levels associated to the topological edge states of
the SSH chains.
These edge states have energies $E_{+} \gtrsim 0$ and $E_{-} \lesssim 0$ \cite{Janos}. More precisely,
 \begin{equation}
 E_{\pm} \propto \pm |v e^{-N_{T}\xi}|,
 \end{equation}
where  $\xi = (\log|w| -\log|v|)^{-1}$  is the localization length of the states $|E_{\pm}\rangle$.
In the limit $v\to 0$, these eigenstates tend to the superpositions of localized edge states given by
\begin{equation}
|E_\pm \rangle =\frac{1}{\sqrt{2}}\left(e^{-i\varphi/2 }|1_{L},0_{R} \rangle \pm e^{i\varphi/2 }|0_{L},1_{R} \rangle \right),
\end{equation}
which are maximally entangled states.\\
In half-filling, for $|\Phi/\Phi_0|< 0.25$ and $\lambda =1^-$,
the state $|E_-\rangle$ is occupied and the state $|E_+\rangle$ is empty. Consequently, there is only one electron shared by the edge sites, and
the concurrence $\mathcal{C}_{L,R}$ reaches the maximum value $1$.
When the magnetic flux increases, the energy of one of the states of the QR drops. For  $\Phi/\Phi_0=0.25$ the energy of the ring's state vanishes, the state hybridizes with the edge states zero modes, and the concurrence $\mathcal{C}_{L,R}$  falls to $1/2$. When $0.25<|\Phi/\Phi_0|<0.5$, the energy of the same
state of the ring drops below zero and absorbs the edge electron. Consequently, both edge levels get empty, and the edge-edge concurrence falls to zero.

\section{Conclusions} \label{conclusions}
In this work we investigated the topological states of a Su-Schrieffer-Heeger (SSH) chain with an embedded Aharonov-Bohm quantum ring. We payed special attention to the states' entanglement-related features, focusing on the entanglement between fermionic modes of the system.
We have found that, in addition to the topological edge states exhibited by the standard SSH chain, our present system also admits two kinds of topological states in the bulk.
On the one hand,  a topological state with a vanishing energy not depending on the magnetic flux, arises in the region usually known as the trivial zone. On the other hand, bulk topological states
with energies depending on the flux, emerge in the standard topological zone. In both cases the ring plays the role of a domain wall, where it is known that another zero energy topological state also occurs \cite{Janos}.\\

Part of this work was devoted to the analysis of the quantum mode entanglement exhibited by pairs of sites in the system. We found that the effects of the magnetic flux on the entanglement between adjacent sites
are small and, consequently, when considering adjacent sites, the behavior of our system is consistent with the one observed in a single SSH chain \cite{Cho}. On the other hand, the ring's magnetic flux has a strong effect on the of edge-edge entanglement. This effect is determined by the state of occupation of the zero energy eigenstates located at the edges. The concurrence between edges depends on the magnetic flux $\Phi$. When the chain's asymmetry parameter $\lambda$ approaches the limit $\lambda \rightarrow 1^-$, the oscillations of the concurrence, as a function of $\Phi$, tend to reach the extreme values $0$ and $1$. In the limit
$\lambda \rightarrow 1^-$, the dependence of the concurrence on the flux has the shape
of a ``square wave", with the concurrence changing abruptly between $0$ and $1$
when $\Phi/\Phi_0$ approaches the values $l+1/4$ ($l$ integer). The dependence of entanglement on the ring's magnetic flux provides a mechanism that can be used to design an entanglement switch between the system's edges. \\

The present work suggests various possible directions for further research. Here we have focused on the entanglement features of a Su-Schrieffer-Heeger chain with an embedded Aharonov-Bohm ring, related to the eigenstates of the system. It would be interesting to investigate explicit dynamical entanglement features of this system, corresponding to time-dependent states, and compare them to the entanglement dynamics exhibited by other fermionic systems, such as those considered in \cite{MF2016,MF2018}. It would also be worth to explore possible connections between our developments, and very recent works on the topology and/or the entanglement in junctions of free fermionic chains akin to the ones consider by us, such as  \cite{C2022} (where the junction is conformally-invariant),  \cite{F2022} (where the junction is a scatterer), or \cite{A2022} (where the junction is a hopping defect). Any advances along these or related lines will certainly be welcome.


\begin{center}
\appendix{APPENDIX: Many Fermion State, Convention, and Basic Issues}
\end{center}

Here, in order to make our work self-contained, we briefly describe the procedure for calculating the reduced density matrix in the occupation number representation. In this representation, each state of the many fermion basis is labeled by a sequence of zeros and ones. The zeros (ones) of a given sequence corresponds to the empty (occupied) one fermion states.
It is well-known that in this formalism there is an implicit convention related to the anticommutative character of fermionic many-body states. This convention plays a relevant role
in the computation of reduced density matrices.

Let us consider a lattice system with $r$ electrons distributed in $N_L$ sites.
In the occupation number representation, each lattice site can be considered as a physical object with two available states, empty ($0$)  or occupied  ($1$).
A state of the system in this representation can be described as,
\begin{equation}
|I;n_{1},n_{2},\dots,n_{N_{L}} \rangle,
 \label{occupationnumber}
\end{equation}
where $n_s=1$ if site $s$ is occupied and $n_s=0$ otherwise, and the ordered multi-index $I=( i_1<i_2 <\dots <i_r)$, where $1\leq  i_j \leq N_L$ indicates explicitly which sites of the lattice $L$ are occupied.
In this description, each site of the lattice defines a mode $s$, and the value of $n_s$ specifies its  state.
In the standard notation the label $I$ is not present in the ket $|I;n_{1},n_{2},\dots,n_{N_{L}}\rangle$, but here we include it for convenience.
We also assume that each multi-index appearing in what follows is in increasing order, unless otherwise indicated.

In the second quantization formalism, each mode $i$ is associated to creation and annihilation operators, $c_i^\dag$ and $c_i$, respectively.
The operator $c^{\dag}_{i}$ ($c_{i}$) creates (annihilates) a fermion in a site $i$ when this site is empty (occupied), otherwise it returns a null vector. These operators obey the anticommutation relations
\[
\{c_{i},c_{j}\}=\{c_{i}^{\dag},c_{j}^{\dag}\}=0, \;    \{c_{i}
,c_{j}^{\dag}\}=\delta_{ij},
\]
 where $\delta_{ij}$ is the Kronecker delta and the curly brackets mean $\{\hat{O}_1,\hat{O}_2\}=\hat{O}_1\hat{O}_2+\hat{O}_2\hat{O}_1$ for any pair of operators $\hat{O}_1$ and $\hat{O}_2$.

The vacuum state $|\emptyset\rangle$ corresponds to a state with $n_s=0$ for all $1\leq  s \leq N_L$.
The state $|I;n_{1},n_{2},\dots,n_{N_{L}}\rangle$ is obtained by the action of the creation operators $c_{i}^{\dag}$ on the vacuum state $|\emptyset\rangle$ as follows,
\begin{subequations}
\begin{equation}
|I;n_{1},n_{2},\dots,n_{N_{C}}\rangle=\prod_{l=1}^{r}c^{\dag}_{i_{l}}|\emptyset \rangle,
\label{occupation-creation}
\end{equation}
where we adopt the convention,
\begin{equation}
\prod_{l=1}^{r}c^{\dag}_{i_{l}}=c^{\dag}_{i_{r}}\dots c^{\dag}_{i_{2}}c^{\dag}_{i_{1}}, \quad i_{1}<i_{2}<\dots<i_{r}. \label{product_}
\end{equation}
\label{fromvacuum}
\end{subequations}
Note that in the definition of the state in the occupation number representation via  Eqs. (\ref{fromvacuum}) it was implicitly adopted a convention of sign.
If it is carried out  an odd permutation of the order in which the creation operators are applied,  a change of sign in the resulting state must be included.
Whatever is the sign convention chosen, it must be applied systematically when calculating quantum correlations between fermion modes.

If $J=(j_{1},\dots,j_{r})$ is a permutation of
$I=(i_{1},\dots,i_{r})$ then,
 \[
|J; n_1,\dots n_{N_L}\rangle =  \prod_{l=1}^{r}c^{\dag}_{j_{l}}|\emptyset \rangle =   \delta^{J}_{I} |I; n_1,\dots n_{N_L}\rangle,
 \]
where $\delta_{I}^{J}$ stands for the
generalized Kronecker delta symbol (see (\ref{generalkronecker})) and
$I$ and $J$ are not necessarily in increasing order.\\
Let $A|B$ be a partition of the lattice $L$, i.e., $L=A\cup B$ and $A\cap B=\emptyset$, given by,
\begin{subequations}
\begin{eqnarray}
 A&=&\{\alpha_{1},\dots,\alpha_{N_{A}}\} \\
 B&=&\{\beta_{1},\dots,\beta_{N_{B}}\}
\end{eqnarray}
\end{subequations}
where $N_{A}+N_{B}=N_{L}$.
Let us consider a state of $p$ fermions distributed among the sites represented by ${J}=(j_{1}<j_{2}<\dots<j_{p})$ of subpart $A$, and another of $q$ fermions  distributed among $K=(k_{1}<k_{2}<\dots<k_{q})$ sites of subset $B$, with $p+q=r$, the number of fermions in the whole lattice. These states can be written as,
\begin{subequations}
\begin{eqnarray}
|J;n_{\alpha_{1}},\dots,n_{\alpha_{N_{A}}}\rangle&=&\prod_{l=1}^{p}c^{\dag}_{j_{l}}|\emptyset\rangle ,\\
|K;n_{\beta_{1}},\dots,n_{\beta_{N_{B}}}\rangle&=&\prod_{l=1}^{q}c^{\dag}_{k_{l}}|\emptyset\rangle,
\end{eqnarray}
\end{subequations}
The wedge product between $|J;n_{\alpha_{1}},\dots,n_{\alpha_{N_{A}}}\rangle$ and $|K;n_{\beta_{1}},\dots,n_{\beta_{N_{B}}}\rangle$ is defined as,
\begin{equation}
|J;n_{\alpha_{1}},\dots,n_{\alpha_{N_{A}}}\rangle\wedge |K;n_{\beta_{1}},\dots,n_{\beta_{N_{B}}}\rangle=\prod_{l=1}^{p}c^{\dag}_{j_{l}}\prod_{l=1}^{q}c^{\dag}_{k_{l}}|\emptyset\rangle.
\end{equation}
This product will allow us later  to obtain  the reduced density matrix of the subparts $A$ or $B$ in the fermionic modes representation.

Given a bipartite system $A|B$ in a state described by a density matrix $\rho$,  the reduced density matrix of the subpart $A$ is obtained by tracing out the degrees of freedom of the subpart $B$,
\begin{equation}
\rho_A = \mathrm{Tr}_B \rho.
\end{equation}
Physically, the reduced  density operator $\rho_{A}$ ($\rho_{B}$) stores the available information of the state $\rho$  located in subpart  $A$ ($B$) ignoring any information associated to subpart $B$ ($A$) \cite{Fermionic-mode}.
The reduced density matrix $\rho_A$ has to obey the consistency condition,
\begin{equation}
  \mathrm{Tr}_{A}(\rho_{A}O_{A}) =   \mathrm{Tr}(\rho O_{A}),\label{consistencia2}
\end{equation}
where $O_{A}$ an observable acting over subpart $A$. Besides, the trace and partial trace
operations have to comply with linearity,
\begin{subequations}
\begin{eqnarray}
  \mathrm{Tr}(\mathrm{a}_{1}\rho^{(1)}+\mathrm{a}_{2}\rho^{(2)}) &=& \mathrm{a}_{1} \mathrm{Tr}( \rho^{(1)})+\mathrm{a}_{2} \mathrm{Tr}(\rho^{(2)}),\nonumber \\
 \label{linearidad}   \\
\mathrm{Tr}_{B}(\mathrm{a}_{1}\rho^{(1)}+\mathrm{a}_{2}\rho^{(2)}) &=& \mathrm{a}_{1} \mathrm{Tr}_{B}(\rho^{(1)})+\mathrm{a}_{2}\mathrm{Tr}_{B}(\rho^{(2)}),\nonumber
\\
\label{linearidad-reducida}
\end{eqnarray}
\label{linear}
\end{subequations}
where  $\mathrm{a}_{i}$  ($i=1,2$) are real numbers. The operators $\rho^{(i)}$ ($i=1,2$) are density matrices of the whole system.

The antisymmetric character of fermionic many-body states, in addition to the  requirements
imposed by (\ref{consistencia2}) and  (\ref{linear}), completely determine the
expressions for the trace $ \mathrm{Tr}(\,)$ and the partial trace $\mathrm{Tr}_{A}(\,)$  ($\mathrm{Tr}_{B}(\,)$) operations. We shall focus on $ \mathrm{Tr}_{B}(\,)$, leaving aside $ \mathrm{Tr}_{A}(\,)$, since it obviously has the same form as the former.

Returning to the our  lattice $L$  and its partition $A|B$,  let us suppose that $I=(i_{1}<i_{2}<\dots < i_{p+q})$  results from a permutation $\sigma$ of the set of numbers $JK=(j_{1}<j_{2}<\dots <j_{p},k_{1}<k_{2}<\dots< k_{q})$  (note that in general $JK$ is not in increasing order). Hence
\begin{eqnarray}
|J;n_{\alpha_{1}},\dots,n_{\alpha_{N_{A}}}\rangle\wedge |K;n_{\beta_{1}},\dots,n_{\beta_{N_{B}}}\rangle\nonumber\\=\delta_{JK}^{I}|I;n_{1},\dots,n_{N_{L}}\rangle.
\end{eqnarray}
Analogously, the wedge product between the bras $\langle J;n_{\alpha_{1}},\dots,n_{\alpha_{N_{A}}}|$ and $\langle K;n_{\beta_{1}},\dots,n_{\beta_{N_{B}}}|$ results in,
\begin{eqnarray}
\langle J;n_{\alpha_{1}},\dots,n_{\alpha_{N_{A}}}|\wedge \langle K;n_{\beta_{1}},\dots,n_{\beta_{N_{B}}}|\nonumber\\=\delta_{JK}^{I}\langle I;n_{1},\dots,n_{N_{L}}|.
\end{eqnarray}
Now we are in position to include into the trace $\mathrm{Tr}_{B}(\,)$ of operator $|I;n_{1},n_{2},\dots,n_{N_{L}}\rangle\langle I'; n_{1},n_{2},\dots,n_{N_{L}}|$ the underlying many fermion antisymmetric property,
\begin{eqnarray}
\mathrm{Tr}_{B}\big(|I;n_{1},n_{2},\dots,n_{N_{L}}\rangle\langle I'; n_{1},n_{2},\dots,n_{N_{L}}|\big)=\nonumber\\ \delta_{JK}^{I}\delta_{J'K'}^{I'}\delta_{K,K'}|J;n_{\alpha_{1}},\dots,n_{\alpha_{N_{A}}}\rangle\langle J';n_{\alpha_{1}},\dots,n_{\alpha_{N_{A}}}|\nonumber\\\label{reduced-operator}
\end{eqnarray}
where as usual $\delta_{K,K'}=1$ if $K=K'$ and $\delta_{K,K'}=0$ otherwise. In the above equations we have used $I,J,K$ ($I',J',K'$) to label kets (bras).
With this procedure, the degrees of freedom associated to the fermion modes belonging to subpart $B$ are eliminated, obtaining the reduced operator of the subpart $A$ in the fermionic modes context.\\
Now, let us consider the density operator,
\begin{equation}
\rho=\sum_{I,I'}[\rho]_{I,I'}|I;n_{1},\dots,n_{N_{L}}\rangle\langle I'; n_{1},\dots,n_{N_{L}}|,
\end{equation}where $ [\rho]_{I,I'}$ are complex numbers, which by definition satisfy the following properties,
\begin{eqnarray}
  [\rho]_{I,I'} =[\rho]^{*}_{I',I}  \\
\mathrm{Tr}(\rho) = \sum_{I}[\rho]_{I,I}= 1
\end{eqnarray}
where the symbol "$*$" denotes complex conjugation.
Now, to calculate the reduced fermion density operator $\rho_{A}$ we apply Eq. (\ref{linearidad-reducida}), i.e., the linearity of partial trace condition,
\begin{eqnarray}
\rho_{A}&=&\mathrm{Tr}_{B}(\rho)\nonumber \\
&=&\sum_{I,I'}[\rho]_{II'}\mathrm{Tr}_B\big(|I;n_{1},\dots,n_{N_{L}}\rangle\langle I'; n_{1},\dots,n_{N_{L}}|\big), \nonumber \\
\label{reduced-rho-A1}
\end{eqnarray}
what results in,
\begin{equation}
\rho_{A}=\sum_{J,J'}[\rho_{A}]_{J,J'}|J;n_{\alpha_{1}},\dots,n_{N_{A}}\rangle\langle J'; n_{\alpha_{1}},\dots,n_{N_{A}}|.\label{reduced-rho-A2}
\end{equation}
From Eqs. (\ref{reduced-operator}), (\ref{reduced-rho-A1}) and (\ref{reduced-rho-A2}) it follows that the coefficients $[\rho_{A}]_{J,J'}$ of reduced density matrix  can be written as,
\begin{equation}
[\rho_{A}]_{J,J'}=\sum_{I,I',K}\delta_{JK}^{I}\delta_{J'K}^{I'}[\rho]_{I,I'}.\label{densidadfermion}
\end{equation}

\providecommand*{\mcitethebibliography}{\thebibliography}
\csname @ifundefined\endcsname{endmcitethebibliography}
{\let\endmcitethebibliography\endthebibliography}{}

\end{document}